
\input phyzzx
\quad\quad\quad\quad\quad\quad\quad\quad \quad\quad
\quad\quad\quad\quad\quad\quad\quad\quad\quad\quad\quad
\quad\quad\quad\quad\quad\ BIR/PH/92-1.
\vskip 1.5cm
\centerline{\bf{Restricted Supergauge invariance,}}
\centerline{\bf{N=2 Coadjoint Orbits}}
\centerline{\bf{and N=2 Quantum Supergravity}}
\vskip 3.5cm
\centerline {W.A. Sabra}
\centerline{Physics Department}
\centerline{Birkbeck College}
\centerline{University of London}
\centerline{Malet Street}
\centerline{London WC1E 7HX}
\vskip 3cm
\REF\poly{A.M. Polyakov, Mod. Phys. Lett. A2 (1987) 893; \hfill\break
V.G. Knizhnik, A.M. Polyakov and A.B. Zamolodchikov, Mod. Phys.
Lett. A3 (1988) 819; \hfill\break
A.M. Polyakov and A.B. Zamolodchikov, Mod.  Phys. Lett. A3
(1988)1213;\hfill\break
Lectures given by Polyakov at Les Houches summer school on Fields,
Strings and Critical Phenomena (1988).}
\REF\gri{ M.T. Grisaru and R.M. Xu, Phys.
Lett. 205B (1988) 486.}
\REF\waf {W. A. Sabra, Int. J. of
Mod. Phys, A 6, 755 (1991)}
\REF\xu{R. Xu, $Two-Dimensional\ Quantum\ (0,2)\ Supergravity$,
preprint UTTG-09-90}
\REF\kura{T. Kuramoto, Nucl. Phys B346 (1990) 527.}
\REF\diff{ A. M. Polyakov, Int. J. Mod. Phys. A5(1990) 833.}
 \REF\Alek{A. Alekseev and Shatashvilli, Nucl. Phys B323
(1989) 719.}.  \REF\oog { M. Bershadsky and H. Ooguri Comm. Math.
Phys.  2(1989)49.}
\REF\wess{J. Wess and B. Zumino, Phys. Lett. 37B (1971) 95.}
\REF\Novi{S. P. Novikov, Sov. Math. Dokl. 24 (1981) 222.}
\REF\WIT{E. Witten, Comm. Math. Phys. 92 (1984) 455.}
\REF\composition{A.M. Polyakov and P.B. Wiegmann, Phys. Lett. 141B
(1984) 233.}
\REF\supoog{ M. Bershadsky and H. Ooguri, Mod. Phys. Lett, 229B (1989) 374.}
\REF\superwaf{ W. A. Sabra,
$\ Hidden \ Kac-Moody\ symmetry\ and\ 2D\ Quantum$\hfill\break
$Supergravity$, to appear in Nucl. phys B}
\REF\ROD{G. Delius, P. van Nieuwenhuizen
and V.G.J. Rodgers,  Int. J. Mod. Phys. A5 (1990) 3943.}
\REF\KIR{Elements of the Theory
of  Representations, A.A. Kirillov, Springer Verlag (1976).}
\REF\wit{E. Witten, Comm. Math. Phys. 114 (1988) 1.}
\REF\gates{R. Brooks, F. Muhammad and S.J.
Gates, Nucl. Phys. B261 (1986) 599.} \REF\N{W. Boucher, D. Friedan and A. Kent,
 Phys. Lett. 172B
(1986) 316 ; S. Nam, Phys. Lett. 172B (1986) 323 ; P. Di Vecchia,
J.L. Petersen, M. Yu and H. B. Zhug, Phys. Lett. 174B (1986) 280 ;
A.B. Zamolodchikov and V.A. Fateev, Zh. Eksp. Theor. Fiz. 90 (1986)
1553.}
\REF\Kazama{V.G. Kac and T. Todorov, Comm. Math. Phys. 102 (1985)
337; Y. Kazama and H. Suzuki, Nucl. Phys. B321 (1989)
232.}
\REF\cohen{ J. D. Cohen, Nucl. Phys. B284 (1987) 349.}
\REF\waki{M. Wakimoto, Comm. Math. Phys. 104 (1986)
605.}\REF\unpublished{A.B. Zamolodchikov. unpublished}

\centerline{ABSTRACT}
It is shown that the N=2 superconformal transformations are
restricted  N=1 supergauge transformations of
a supergauge theory with Osp(2,2) as a gauge group. Based on this
result, a canonical derivation of the Osp(2,2) current algebra in
the superchiral gauge formulation of N=2 supergravity is presented.
\vfill\eject
\section {Introduction}The importance of two
dimensional $(2d)$ gravity and supergravity is crucial in providing
an understanding of non-critical string theories and lattice models
formulated on random surfaces. In the continuum formulation of
these models, it became evident from the work of Polyakov et al.
[\poly], that a further insight into the quantization of induced
$2d$ quantum gravity is obtained if one employs a particular
gauge, known as the chiral gauge.  This particular choice of gauge
led Polyakov to discover that the theory possesses  a gauge symmetry
based upon the non-compact group SL(2,R), a result which was
obtained by an explicit calculation of the correlation functions for
the gravitational field surviving the chiral gauge. The extension
of these results to the cases of both  N=1 and N=2 $2d$ supergravity
has also been performed [\poly,\gri,\waf,\xu,\kura].

It has  now  been established that the appearance of the
SL(2,R) symmetry in induced $2d$ quantum gravity is
connected to the fact that the structure of conformal symmetry
exhibits a hidden SL(2,R) current algebra symmetry
[\diff,\Alek,\oog].

In [\diff], it was demonstrated that diffeomorphisms
can be obtained from restricted SL(2,R) gauge transformations.
One starts with a two dimensional gauge theory described by
the gauge fields $A_za$ and $A_{\bar z}a$ \footnote *{we
parametrize the two dimensional space time with coordinates $(x, t)$
by $z=t+x$ and $\bar z=t-x$.} where $a$ takes values in
the set $\{+,-,0\}$ and is the SL(2,R) group index. Now
partially fix a gauge by imposing the three conditions,  $$A_z+=1,
\qquad A_z0=0,\qquad A_z{-}=T.\eqn\jhj$$ In this gauge the
residual gauge transformation of $T$ becomes the action of Virasoro
algebra on the spin-2 stress energy tensor and therefore  the
dynamics of the restricted gauge theory describes the geometric
quantization of the Virasoro algebra [\wit]. The reason why the gauge
field $A_z-$ becomes a spin two field is because in the \lq\lq
background field" $A_z{+}=1,$ the internal isotopic space becomes
equivalent to a two dimensional space-time [\diff].  This geometrical
observation can then be employed to explain the relationship between
the Wess-Zumino-Novikov-Witten (WZNW) action
[\wess,\Novi,\WIT,\composition] with the group SL(2,R) and that of
Polyakov's two dimensional gravitation action. An equivalent
analysis has been performed in [\Alek,\oog] using the method of
Hamiltonian reduction.

In [\superwaf] the results of [\diff] were generalized to the
case of N=1 superconformal symmetry. It was shown that  N=1
superdiffeomorphisms can be obtained from restricted N=1 Osp(1,2)
supergauge tranformations. By exploiting this, the relationship
existing between induced N=1 $2d$ supergravity and  N=1
Osp(1,2) WZNW is analysed.

In this paper, the case of N=2 superconformal symmetry
will be considered. It  will be shown that the N=2 superconformal
symmetry [\N] can be obtained from restricting the supergauge
transformations of a $(1,0)$ supergauge theory with Osp(2,2) as a
gauge group. Based on this result, the method of
[\diff,\superwaf] is generalized to the case of induced $(2,0)$
$2d$ supergravity. We also comment on the case of induced $(2,2)$
$2d$ supergravity. The case of N=2 $2d$ supergravity  is of
particular interest since unlike the case of N=0 and N=1
$2d$ supergravity,  its chiral gauge formulation is valid for any
space-time dimension. Thus it is relevant to the study of
supersymmetric four dimensional noncritical strings.

This work is organised as follows. In section 2,  the
formulation of induced  $(2,0)$ $2d$ supergravity in the superchiral
gauge [\waf,\xu] is reviewed.  The relationship between the action
constructed on the coadjoint orbit (of purely central extension) of
the N=2 superVirasoro group [\ROD] and that of induced $(2,0)$
$2d$ supergravity in the superchiral gauge is also presented. In
section 3, a $(1,0)$ supergauge theory with the gauge group Osp(2,2)
is considered. By fixing a partial gauge, a reduced theory
is obtained which has the N=2 superconformal symmetry. The
relationship between the induced $(2,0)$ $2d$ supergravity action in
the superchiral gauge formulation and the geometric action of N=2
superVirasoro group is then derived, providing a canonical
derivation of the Osp(2,2) current algebra symmetry in induced
$(2,0)$ $2d$ supergravity. In addition, a composition formula for the
geometric action of the N=2 superconformal group is also derived.
This is the $(2,0)$ supergravitational analogue of the
Polyakov-Wiegmann identity of the WZNW model [\composition]. In the
last section, we discuss our results and suggest a relationship
linking, respectively, the super current algebras Osp(1,2) and
Osp(2,2) with the N=1, 2 superconformal algebras.

\section{ $(2,0)$
supergravity in the superchiral gauge}  In this section, the
formulation of induced $(2,0)$ $2d$ supergravity in the superchiral
gauge is  reviewed. The $(2,0)$ superspace is described by two
Grassmann coordinates $(\theta+_1,\theta_2+),$ which can be
combined into a single complex Grassmann coordinate, $\theta.$ The
$(2,0)$ superspace  coordinates  are thus given by the set $(z,\bar
z,\theta,\bar\theta)$. The rigid $(2,0)$ supersymmetry algebra can
then be described as, $$\{D_\theta,
D_{\bar\theta}\}=2\partial_z,\quad \{D_\theta,D_\theta\}=\{
D_{\bar\theta},
D_{\bar\theta}\}=[D_\theta,\partial_z]=[D_\theta,\partial_{\bar
z}]=0,\eqn\super$$ where
$D_\theta=\partial_\theta+\bar\theta\partial_z$,
$D_{\bar\theta}=\partial_{\bar\theta}+\theta\partial_z.$  In curved
superspace, the supercovariant derivatives are given by,
$$\nabla_A=EM_AD_M+w_AM,\eqn\fibre$$ where $E_AM$ are the
vielbeins, $w_A$ are the spin connections and $M$ is the Lorentz
generator. The constraints in $(2,0)$ $2d$ supergravity [\gates] are
$$\eqalign{\{\nabla_\theta,\nabla_{\bar\theta}\}=&2\nabla_z,\quad
\{\nabla_\theta,\nabla_\theta\}=
\{\nabla_{\bar\theta},\nabla_{\bar\theta}\}=0,\cr
[\nabla_\theta,\nabla_{\bar z}]=&iG_{\bar
z}D_\theta+2{\Sigma}{\bar\theta} M,\cr [\nabla_{z},\nabla_{\bar
z}]=&{\Sigma}\theta
\nabla_\theta+{\Sigma}{\bar\theta}\nabla_{\bar\theta}+RM,\cr}\eqn\grav$$
where  $\Sigma{\bar\theta}$ and $R$ are the superfields  whose
first components are the supercovariant field strengths for the
component gravitino and graviton respectively.  In order for the
constraints to satisfy the Bianchi identities, the following
relations must hold,
$$\eqalign{\nabla_{\theta}G_{\bar
z}=&\Sigma{\bar\theta},\quad\nabla_{\theta}\Sigma{\bar\theta}=0,
\cr\nabla_{\theta}\Sigma{\theta}+
\nabla_{\bar\theta}\Sigma{\bar\theta}&=R,\quad
\nabla_{\theta}R=2\nabla_{z}\Sigma{\bar\theta}.\cr}\eqn\wrinkles$$
In solving the constraints and choosing the superchiral gauge, it
was found [\waf,\xu] that the theory can only be described in terms
of the unconstrained superfield $H_{\bar z\bar z}$ and that the
solution of the constraints is given
by$$\eqalign{\nabla_\theta=&D_\theta,\quad \nabla_{\bar\theta}=
D_{\bar\theta},\quad \nabla_z=\partial_z,\cr\nabla_{\bar
z}=&\partial_{\bar z}+{1\over2}(D_\theta H_{\bar z\bar
z})D_{\bar\theta}+{1\over2}(D_{\bar\theta} H_{\bar z\bar z})
D_\theta+H_{\bar z\bar z} \partial_z +\partial_z H_{\bar z\bar z}
M,\cr  R=&\partial2_z H_{\bar z\bar z}, \quad
\Sigma{\bar\theta}={1\over2}\partial_z D_\theta H_{\bar z\bar
z},\quad G_{\bar z} ={1\over 4i}[D_\theta,D_{\bar\theta}]H_{\bar
z\bar z}.\cr}\eqn\aoun$$  The equation of motion of the superfield
$H_{\bar z\bar z}$ is derived by using the anomaly equation
[\waf,\xu],  $$\partial_z [D_\theta, D_{\bar\theta}]H_{\bar z\bar
z}=0. \eqn\anom$$  Expanding $H_{\bar z\bar z}$ in terms of a set
of fields $Ja$ as, $$H_{\bar z\bar z}=z2 J{-1}-2z J0+J1-\theta z
J{-\bar{1\over2}}-\bar\theta z J{-{1\over2}} +\theta
J{\bar{1\over2}}+\bar\theta J{1\over2}+i\theta\bar\theta
J{U(1)},\eqn\invasion$$  and then calculating the Ward identities
of the theory in terms of the fields $Ja$, it was demonstrated
[\waf,\xu] that the theory  possesses an associated Osp(2,2) current
algebra.

Before discussing the relation of $(2,0)$ $2d$ supergravity in the
superchiral gauge to the coadjoint orbits of the N=2 superVirasoro
group, we briefly review the construction of dynamical systems
having the coadjoint orbit [\KIR,\wit] of a Lie group as a phase
space.  On the coadjoint orbits of a Lie group G, a G-invariant
symplectic structure can be defined [\KIR], that is, there exists a
natural antisymmetric bilinear form which is both closed and
nondegenerate. This symplectic structure can be constructed as
follows. An element $u$ of the Lie algebra ${\cal G}$ of G maps a
coadjoint vector $a$ of the smooth dual space ${\cal G}*$ to the
coadjoint vector $u(a)$ defined by    $$(u(a))(v)=-a([u,v]); \qquad
\forall v\in {\cal G}, \quad a\in {\cal G}*.\eqn\egypt$$
Fix a covector $b$ and represent by $W_b$,
the orbit of $b$ obtained  by the action of G on $b$.  Let $a$ and
$a'$ be  coadjoint vectors in $W_b,$ tangent to the orbit at $b$,
being reached by  applying an infinitesimal group transformation at
$b$. Thus, there exist two vectors $u$ and $u'$ satisfying,
$$u(b)=a;\qquad u'(b)=a'.\eqn\jose$$ The symplectic 2-form $\omega$,
is given by [\KIR],
$$\omega(a,a')=b([u,u']).\eqn\magic$$
 The action $S$ of a dynamical system defined on $W_b$  can now be
constructed by integrating the two form $\omega$ over a two
dimensional submanifold $\Sigma$ of the coadjoint orbit,
$$S=\int_{\Sigma} b([u,u']).\eqn\sowhat$$The above algorithm has
been used to construct actions on the  coadjoint orbits of the
Kac-Moody and Virasoro groups and their supersymmetric  extensions.
(For a review see [\ROD] and references therein.)

In the case of the N=2 superVirasoro group, the elements of the
group are given by the superdiffeomorphisms $X$ and $\Theta,$
satisfying the chirality and the superconformal conditions,
$$\eqalign{&D_\theta\Theta=0,\qquad
D_{\bar\theta}\bar\Theta=0\qquad D_{\theta}X=\Theta
D_{\theta}\bar\Theta, \qquad D_{\bar\theta}X=\bar\Theta
D_{\bar\theta}\Theta.\cr}\eqn\hell$$ The action
constructed on the coadjoint orbit of purely central extension is
given by  $$S{(2,0)}_{s.vir}=\int d2zd2\theta\Big(
{\partial_z\Theta\partial_{\bar z}\bar\Theta-
\partial_z\bar\Theta\partial_{\bar z}\Theta\over
(D_{\bar\theta}\Theta)(D_{\theta}\bar\Theta)}\Big).\eqn\telephone$$

The relation of the induced $(2,0)$ $2d$ supergravity
action, when formulated in the superchiral gauge to \telephone\
becomes transparent if one parametrizes the superfield  $H_{\bar z
\bar z}$ as $$H_{\bar z \bar z}={\partial_{\bar
z}f+\bar\psi\partial_{\bar z}\psi+ \psi\partial_{\bar
z}\bar\psi\over\partial_{ z}f+\bar\psi\partial_{z}\psi+
\psi\partial_{z}\bar\psi},\eqn\martin$$ where $f$ and $\psi$ are
respectively Bose and Fermi $(2,0)$ superfields satisfying the same
conditions as $X$ and $\Theta,$ i.e.,
$$\eqalign{&D_\theta\psi=0,\qquad
D_{\bar\theta}\bar\psi=0,\qquad D_{\theta}f=\psi
D_{\theta}\bar\psi, \qquad D_{\bar\theta}f=\bar\psi
D_{\bar\theta}\psi.\cr}\eqn\he$$
Under the infinitesimal $(2,0)$ superdiffeomorphisms,
$$z\rightarrow z+\delta z, \quad \theta\rightarrow
\theta+\delta\theta, \quad \bar\theta\rightarrow
\bar\theta+\delta\bar\theta,\eqn\guildford$$
the transformation of the new superfields are given by
$$\eqalign{&\delta\psi={\cal
E}z\partial_z\psi+{1\over2}D_\theta{\cal E}z
D_{\bar\theta}\psi,\cr
 &\delta\bar\psi={\cal E}z
\partial_z\bar\psi+{1\over2}D_{\bar\theta}{\cal E}z
D_{\theta}\bar\psi,\cr &\delta f={\cal E}z\partial_z
f+{1\over2}D_\theta{\cal E}z D_{\bar\theta}f+
{1\over2}D_{\bar\theta}{\cal E}zD_{\theta}f.\cr}\eqn\she$$
where ${\cal E}z=\delta
z+\theta\delta\bar\theta+\bar\theta\delta\theta.$ Using the
relations \martin, \he\ and \she, one can recover the transformation
of the superfield $H_{\bar z \bar z}$, which is given by,
$$\delta H_{\bar z \bar z} =\partial_{\bar z}{\cal E}z+
{\cal E}z\partial_zH_{\bar z \bar z}+ {1\over2}D_\theta{\cal
E}zD_{\bar\theta}H_{\bar z \bar z}+ {1\over2}D_{\bar\theta}{\cal
E}zD_{\theta}H_{\bar z \bar z}- \partial_z {\cal E}zH_{\bar z \bar
z}.\eqn\sabra$$ The induced $(2,0)$ $2d$ supergravity, can then be
obtained from \telephone\ via the following set of
transformations,$$X(f,\bar z,\psi,\bar\psi)=z,\quad \Theta(f,\bar
z,\psi,\bar\psi)=\theta, \quad \bar\Theta(f,\bar
z,\psi,\bar\psi)=\bar\theta.\eqn\nestor$$   \section{super gauge
transformations and superdiffeomorphisms} In this section, we show
that the action describing the dynamics of the N=2 superconformal
algebra, i.e., the action \telephone, can be obtained by partially
fixing a certain gauge in the $(1,0)$ supersymmetric Osp(2,2) gauge
theory.  The relationship between the $(2,0)$ $2d$ supergravity
action to the coadjoint action is a consequence of the properties of
the super WZNW  model.

The orthosymplectic group Osp(2,2)  is generated
by four bosonic generators $\{l_0, l_{-1}, l_{1}, l_u\} $ and four
fermionic generators  $\{({l_{1\over2})}_1, {(l_{1\over2})}_2,
{(l_{-{1\over2}})}_1, {(l_{-{1\over2}})}_2\},$ which are represented
as follows, $$l_0=\pmatrix{{1/2}&0&0&0\cr 0&-{1/2}&0&0\cr
0&0&0&0\cr  0&0&0&0\cr}\quad l_{1}=\pmatrix{0&1&0&0\cr 0&0&0&0\cr
0&0&0&0\cr  0&0&0&0\cr}\quad l_{-1}=\pmatrix{0&0&0&0\cr 1&0&0&0\cr
0&0&0&0\cr  0&0&0&0\cr}$$ $$l_u=\pmatrix{0&0&0&0\cr 0&0&0&0\cr
0&0&0&1\cr 0&0&- 1&0\cr}\quad
{(l_{{1\over2}})}_1=\pmatrix{0&0&1&0\cr 0&0&0&0\cr 0&-1&0&0\cr
0&0&0&0\cr}\quad {(l_{{1\over2}})}_2=\pmatrix{0&0&0&1\cr 0&0&0&0\cr
0&0&0&0\cr 0&-1&0&0\cr}$$
$${(l_{-{{1\over2}}})}_1=\pmatrix{0&0&0&0\cr 0&0&1&0\cr 1&0&0&0\cr
0&0&0&0\cr}\quad {(l_{-{{1\over2}}})}_2=\pmatrix{0&0&0&0\cr
0&0&0&1\cr  0&0&0&0\cr 1&0&0&0\cr}.\eqn\christ$$
We consider a $(1,0)$ supersymmetric two dimensional gauge
theory with Osp(2,2) as a gauge group.
The $(1,0)$ superspace is described by the coordinates
$(z,\bar z,\theta+)$ where $\theta+$ is a real Grassmann
coordinates. The rigid $(1,0)$ supersymmetry
algebra can then be described as, $$\{D_+,D_+\}=2\partial_z,\quad
[D_+,\partial_z]=[D_+,\partial_{\bar
z}]=0,\eqn\su$$ where $D_+=\partial_{\theta+}+\theta+\partial_z.$
The $(1,0)$ supergauge theory has two sectors, a supersymmetric
left-moving sector described by the chiral spinor $
A_{\theta+}$ and a right bosonic sector described by the chiral
vector $A_{\bar z}$. The gauge transformations of the gauge fields
are given by, $$\eqalign{\delta A_{\theta+}a=&{\cal
D}_{\theta+}\epsilona=
D_{\theta+}\epsilona-fa_{bc}A_{\theta+}b\epsilonc,\cr\delta
A_{\bar z}a=&{\cal D}_{\bar z}\epsilona=\partial_{\bar
z}\epsilona-fa_{bc}A_{\bar z}b\epsilonc,\cr}\eqn\spain$$where
$fa_{bc}$ are the structure constants of the Osp(2,2) algebra and
$\epsilona$ are the gauge parameters. The effective action of the
gauge field $A_{\theta+}$ is given by,  $$S(A_{\theta+})\sim
\hbox{log}\ \hbox{sdet} (D_{\theta+}-A_{\theta+}).\eqn\leo$$   Its
variation under gauge transformations is,  $$\delta
S(A_{\theta+})=\int d2zd\theta+\ \hbox{str}(J_{\bar z}\delta
A_{\theta+}),\eqn\leon$$ where $J_{\bar z}$ is the gauge current
satisfying the anomaly equation,$${\cal D}_{\theta+}
J_{\bar z}=-k\partial_{\bar z}A_{\theta+},\eqn\leona$$
and $k$ is a constant. Using this anomaly equation together
with the gauge transformations of the gauge fields, we then obtain,
$$\eqalign{\delta S(A_{\theta+})&=\int d2zd{\theta+}
\ \hbox{str}\Big(J_{\bar z}{\cal D}_{{\theta+}}\epsilon\Big)\cr
&=-\int d2zd{\theta+} \ \hbox{str}\Big(\hat\epsilon{\cal
D}_{{\theta+}}J_{\bar z}\Big)\cr &=k\int d2zd{\theta+}
\ \hbox{str}(\hat\epsilon{\partial_{\bar
z}}A_{\theta+}),\cr}\eqn\leonar$$ where $\epsilon$ is a matrix
gauge parameter taking values in the algebra of Osp(2,2) (and
$\hat\epsilon$ is obtained from $\epsilon$ by multiplying  its
fermionic elements by a factor of minus one).
If we parametrize $A_{\theta+}$
by  $$A_{\theta+}=D_{{\theta+}}gg{-1},\qquad g(z,\bar
z,{\theta+})\in \hbox{Osp(2,2)},\eqn\clouds$$ then the action
$S(A_{\theta+})$ is given by a $(1,0)$ WZNW model
$S_1(g) [\Kazama],$ with Osp(2,2) as a gauge group. Similarly one can
parametrize $A_{\bar z}=\partial_{\bar z}hh{-1}$, where $h$ is a
group element of Osp(2,2)  and find that the effective action of the
gauge field $A_{\bar z}$ is also given  by  a $(1,0)$ Osp(2,2) WZNW
model, $S_2(h).$  The final form of the total effective action is
then, $$S_{eff}(g,h)=S_1(g)+S_2(h)-k\int d2zd{\theta+}
\hbox{str}(D_{\theta+}
gg{-1}\partial_{\bar z}hh{-1}),\eqn\hot$$ where the last term
is added to insure gauge invariance.
In terms of the new parameters, a finite gauge transformation on
$A_{{\theta+}}$ and $A_{\bar z}$ is given by, $$g\rightarrow
Ug,\qquad h\rightarrow Uh, \quad U\in {\hbox
{Osp(2,2)}}.\eqn\bruno$$ As the  effective action  is invariant
under this transformation, this implies the following symmetry,
$$S_{eff}(g,h)=S_{eff}(Ug,Uh).\eqn\roncarati$$
If we set $U=h{-1}$ or $U=g{-1}$, we can then deduce
that
$$S_{eff}(g,h)=S_1(h{-1}g)=S_2(g{-1}h),\eqn\madrid$$
and in particular,
$$S_1(h{-1})=S_2(h).\eqn\israel$$
Finally, using \hot, \madrid\ and \israel, we arrive at the $(1,0)$
supersymmetric extension of the Polyakov-Weigmann composition formula
[\composition],
$$S_1(h{-1}g)=S_1(g)+S_1(h{-1})-k\int d2zd{\theta+}
\ \hbox{str}(D_{\theta+}gg{-1}\partial_{\bar
z}hh{-1}).\eqn\police$$

We will now partially fix a gauge  by imposing the following
conditions,
$$\eqalign{A_{\theta+}0=A_{\theta+}1=&
{(A_{\theta+}{-{1\over2}})}_1=
A_{\theta+}u={(A_{\theta+}{1\over2})}_2=0,\qquad
{(A_{\theta+}{1\over2})}_1=1,\cr
& {(A_{\theta+}{-{1\over2}})}_2=\hbox{unfixed},\qquad
A_{\theta+}{-1}=\hbox{unfixed}.\cr}\eqn\sod$$
In explicit components, the gauge transformations of the
supersymmetric left-moving part of the theory are given by,
$$\eqalign{\delta A_{\theta+}{-1}=&D_{\theta+}\epsilon{-1}+
2(A_{\theta+}{-{1\over2}})_2\epsilon_2{-{1\over2}}
+2(A_{\theta+}{-{1\over2}})_1\epsilon_1{-{1\over2}}+
A_{\theta+}{-1}\epsilon0 -A_{\theta+}0\epsilon{-1},\cr
\delta
A_{\theta+}{1}
=&D_{\theta+}\epsilon{1}-2{(A_{\theta+}{{1\over2}})}_1
\epsilon_1{{1\over2}}
-2{(A_{\theta+}{{1\over2}})}_2\epsilon_2{1\over2}
+A_{\theta+}{0}\epsilon{1} -A_{\theta+}{1}\epsilon{0},\cr
\delta A_{\theta+}u =&D_{\theta+}\epsilon{u}+
{(A_{\theta+}{-{1\over2}})}_1\epsilon_2{{1\over2}}
+{(A_{\theta+}{{1\over2}})}_2\epsilon_1{-{1\over2}}
-{(A_{\theta+}{{1\over2}})}_1\epsilon_2{-{1\over2}}
-{(A_{\theta+}{-{1\over2}})}_2\epsilon_1{{1\over2}},\cr
\delta
A_{\theta+}0=&D_{\theta+}\epsilon0+2(A_{\theta+}{1\over2})_2
\epsilon_2{-{1\over2}}
+2(A_{\theta+}{-{1\over2}})_2\epsilon_2{1\over2}+
2(A_{\theta+}{{1\over2}})_1\epsilon_1{-{1\over2}}
+2(A_{\theta+}{-{1\over2}})_1\epsilon_1{{1\over2}} \cr
&+2A_{\theta+}{1}\epsilon{-1}-2A_{\theta+}{-1}\epsilon1, \cr
\delta {(A_{\theta+}{{1\over2}})}_1=&
D_{\theta+}\epsilon_1{1\over2}+
A_{\theta+}{u}\epsilon_2{{1\over2}}
+A_{\theta+}1\epsilon_1{-{1\over2}}
-{(A_{\theta+}{-{1\over2}})}_1\epsilon{1}
-{(A_{\theta+}{{1\over2}})}_2\epsilon{u}
\cr &+{1\over2}A_{\theta+}{0}\epsilon_1{1\over2}
-{1\over2}{(A_{\theta+}{{1\over2}})}_1\epsilon0, \cr\delta
(A_{\theta+}{1\over2})_2=&
D_{\theta+}\epsilon_2{1\over2}-
A_{\theta+}{u}\epsilon_1{{1\over2}}
+A_{\theta+}{1}\epsilon_2{-{1\over2}}
-(A_{\theta+}{-{1\over2}})_2\epsilon{1}
+(A_{\theta+}{{1\over2}})_1\epsilonu
\cr &+{1\over2}A_{\theta+}0\epsilon_2{1\over2}
-{1\over2}(A_{\theta+}{1\over2})_2\epsilon0,\cr\delta
{(A_{\theta+}{-{1\over2}})}_1=&
D_{\theta+}\epsilon_1{-{1\over2}}+
A_{\theta+}u\epsilon_2{-{1\over2}}
+A_{\theta+}{-1}\epsilon_1{1\over2}
-(A_{\theta+}{1\over2})_1\epsilon{-1}
-(A_{\theta+}{-{1\over2}})_2\epsilonu
\cr &-{1\over2}A_{\theta+}0\epsilon_1{-{1\over2}}
+{1\over2}{(A_{\theta+}{-{1\over2}})}_1 \epsilon0,\cr
\delta {(A_{\theta+}{-{1\over2}})}_2
=&D_{\theta+}\epsilon_2{-{1\over2}}-
A_{\theta+}u\epsilon_1{-{1\over2}}
+A_{\theta+}{-1}\epsilon_2{{1\over2}}
-{(A_{\theta+}{{1\over2}})}_2\epsilon{-1}
+{(A_{\theta+}{-{1\over2}})}_1\epsilon{u}
\cr &-{1\over2}A_{\theta+}{0}\epsilon_2{-{1\over2}}
+{1\over2}{(A_{\theta+}{-{1\over2}})}_2 \epsilon0.\cr}\eqn\sid$$

In order that \sod\ be consistent with the gauge transformations
\sid, the gauge parameters must satisfy,
$$\eqalign{\epsilon_1{-{1\over2}}=&-{1\over2}D_{\theta+}\partial_z
\epsilon1-{(A_{\theta+}{-{1\over2}})}_2\epsilon_2{{1\over2}}
+A_{\theta+}{-1}\epsilon{1},\cr
\epsilon_1{{1\over2}}=&{1\over2}D_{\theta+}\epsilon{1},\cr
\epsilon0=&\partial_z\epsilon{1},\cr
\epsilon{u}=&-D_{\theta+}\epsilon_2{1\over2}+{(A_{\theta+}{-{1\over2}})}_2
\epsilon{1},\cr
\epsilon{-1}=&D_{\theta+}\epsilon_1{-{1\over2}}+
A_{\theta+}{-1}\epsilon_1{{1\over2}}
-{(A_{\theta+}{-{1\over2}})}_2\epsilon{u},\cr
\epsilon_2{-{1\over2}}=&
D_{\theta+}\epsilon{u}-{(A_{\theta+}{-{1\over2}})}_2\epsilon_1{{1\over2}}.
\cr}\eqn\hate$$ These equations, when substituted back into the
transformations of $A_{\theta+}{-1}$ and
${(A_{\theta+}{-{1\over2}})}_2$ give,
$$\eqalign{\delta
A_{\theta+}{-1}=&-{1\over2}\partial_z2D_{\theta+}\epsilon1+
\partial_zA_{\theta+}{-1}\epsilon1+{3\over2}A_{\theta+}{-1}
\partial_z\epsilon1
+{1\over2}D_{\theta+} A_{\theta+}{-1}D_{\theta+}\epsilon1\cr &
-2{(A_{\theta+}{-{1\over2}})}_2\partial_z\epsilon_2{1\over2}
-\partial_z{(A_{\theta+}{-{1\over2}})}_2\epsilon_2{1\over2}+
D_{\theta+}{(A_{\theta+}{-{1\over2}})}_2D_{\theta+}\epsilon_2{1\over2},\cr
\delta {(A_{\theta+}{-{1\over2}})}_2=&-\partial_zD_{\theta+}
\epsilon_2{1\over2}
+\partial_z{(A_{\theta+}{-{1\over2}})}_2\epsilon1
+{(A_{\theta+}{-{1\over2}})}_2
\partial_z\epsilon1+A_{\theta+}{-1}\epsilon_2{1\over2}
\cr &-{1\over2}D_{\theta+}
{(A_{\theta+}{-{1\over2}})}_2D_{\theta+}\epsilon1.\cr}\eqn\deniro$$
If we write$$\eqalign{G_1+\theta+ T=&-kA_{\theta+}{-1},\cr
U-\theta+ G_2 =&-k{(A_{\theta+}{-{1\over2}})}_2,\cr
\varepsilonz+\theta+\varepsilon_1{1\over2}=&\epsilon1
,\cr\varepsilon{1\over2}_2+\theta+\varepsilon=&
-2\epsilon{1\over2}_2,\cr}\eqn\kinsky$$ then the transformations
\deniro\ in components, give us the following equations,
$$\eqalign{\delta_{\varepsilonz}T=&
{k\over2}\partial_z3\varepsilonz+ \partial_z
T\varepsilonz+2T\partial_z\varepsilonz,\cr
\delta_{\varepsilon_1{1\over2}}T=& -{1\over2}\partial_z
G_1\varepsilon_1{1\over2}
-{3\over2}G_1\partial_z\varepsilon_1{1\over2},
\cr
\delta_{\varepsilon_2{1\over2}}T=& -{1\over2}\partial_z
G_2\varepsilon_2{1\over2}
-{3\over2}G_2\partial_z\varepsilon_2{1\over2},\cr
\delta_\varepsilon
T=&U\partial_z \varepsilon.\cr}\eqn\ff$$

$$\eqalign{\delta_{\varepsilonz}G_1=&
\varepsilonz\partial_z G_1+{3\over2}G_1\partial_z\varepsilonz,
\cr
\delta_{\varepsilon_1{1\over2}}G_1=&
{k\over2}\partial2_z\varepsilon_1{1\over2}+
{1\over2}\varepsilon_1{1\over2}T,
\cr
\delta_{\varepsilon_2{1\over2}}G_1=&
+{1\over2}\varepsilon_2{1\over2}\partial_zU
+U\partial_z\varepsilon_2{1\over2},\cr
\delta_\varepsilon G_1=&+{1\over2}\varepsilon G_2,\cr}\eqn\fff$$
$$\eqalign{\delta_{\varepsilonz}G_2=& \varepsilonz\partial_z
G_2+{3\over2}G_2\partial_z\varepsilonz,\cr
\delta_{\varepsilon_1{1\over2}}G_2=&
-{1\over2}\varepsilon_1{1\over2}\partial_zU
-U\partial_z\varepsilon_1{1\over2},
\cr
\delta_{\varepsilon_2{1\over2}}G_2=&
{k\over2}\partial2_z\varepsilon_2{1\over2}+
{1\over2}\varepsilon_2{1\over2}T,
\cr\delta_\varepsilon G_2=&
-{1\over2}\varepsilon G_1,\cr}\eqn\ffff$$
$$\eqalign{\delta_{\varepsilonz}U=&\varepsilonz\partial_z
U+U\partial_z\varepsilonz,\cr
\delta_{\varepsilon_1{1\over2}}U=&
{1\over2}G_2\varepsilon_1{1\over2},
\cr\delta_{\varepsilon_2{1\over2}}U=&
-{1\over2}G_1\varepsilon_2{1\over2},\cr
\delta_\varepsilon
U=&-{k\over2}\partial_z\varepsilon.\cr}\eqn\fffff$$ The equations
\ff, \fff, \ffff\ and \fffff\ represent the N=2 infinitesimal
superconformal transformation [\N], where $T$ is the spin-2 current
generating diffeomorphisms, $G_1$ and $G_2$ are the two spin-$3/2$
supercurrents generating the two supersymmetries and $U$ is a spin-1
current generating the $\hat U(1)$ Kac-Moody algebra.
The extra spin acquired by $T,$ $G_1,$
 $G_2$ and $U$ is due to the fact that in the background
${(A_{\theta+}{1\over2})}_1=1$, the isospin $-{1/2}$ is
equivalent to the spin $1/2$.

In the partially fixed theory, $\delta
S(A_{\theta+})$ reduces to $$\delta S(A_{\theta+})=k\int
d2zd{\theta+} \Big(\epsilon1\partial_{\bar
z}A{-1}-2\epsilon{1\over2}_2\partial_{\bar
z}(A_{\theta+}{-{1\over2}})_2\Big).\eqn\war$$  With the
identification \kinsky, Eq. \war\ describes the dynamics of the N=2
stress-energy tensor, ${\cal U}_z$, $$\delta S({\cal U}_{z})=- \int
d2zd2\theta{\cal E}z \partial_{\bar z}{\cal U}_z,\eqn\fel$$ where
$\delta S({\cal U}_{z})$ represents the variation of $S({\cal
U}_{z})$ under the N=2 superconformal transformation, parametrized
by the $(2,0)$ superfield ${\cal E}z.$ In addition, we can also
write $\delta S({\cal U}_{z})$ as, $$\delta S({\cal U}_{z})=\int
d2zd2\theta L_{\bar z\bar z}\delta {\cal U}_{z},\eqn\safia$$ where
$L_{\bar z\bar z}$ is some  function of ${\cal U}_{z}$. Comparing
the above two equations and using  $$\delta {\cal
U}_z=-{k\over4}[D_{\theta}, D_{\bar\theta}]\partial_z{\cal E}z+
{1\over2}D_{\theta}{\cal E}z D_{\bar\theta}{\cal U}_{z} +
{1\over2}D_{\bar\theta}{\cal E}zD_{\theta} {\cal U}_{z}+ \partial_z
{\cal U}_{z}{\cal E}z+ {\cal U}_{z}\partial_z{\cal E}z\eqn\wafic$$
we deduce that $L_{\bar z\bar z}$ must satisfy the following
equation,  $$\Big(\partial_{\bar z}-L_{\bar z\bar z}\partial_z
-{1\over2}(D_{\bar\theta} L_{\bar z\bar
z})D_\theta-{1\over2}(D_\theta L_{\bar z\bar z})D_{\bar\theta}-
(\partial_z  L_{\bar z\bar z})\Big){\cal U}_{z}
=-{k\over4}[D_\theta, D_{\bar\theta}]\partial_z L_{\bar z\bar
z}.\eqn\hik$$ Defining the action $S(H_{\bar z\bar z})$ as the
Legendre transform of  $S({\cal U}_{z})$ [\diff],  its
transformation under superdiffeomorphisms is then given by, $$\delta
S(H_{\bar z\bar z})=\int d2zd2\theta Z_{z}\delta H_{\bar z\bar
z},\eqn\rock$$ where $Z_{z}$ satisfies the following equation,
$$\Big(\partial_{\bar z}-H_{\bar z\bar z}\partial_z
-{1\over2}(D_{\bar\theta} H_{\bar z\bar
z})D_\theta-{1\over2}(D_\theta H_{\bar z\bar z})D_{\bar\theta}-
(\partial_z  H_{\bar z\bar z})\Big)Z_{z} =-{k\over4}[D_\theta,
D_{\bar\theta}]\partial_z H_{\bar z\bar z},\eqn\Gabo$$ and the
transformation of the superfield $H_{\bar z\bar z}$ is given by
\sabra.   Finally, we define the combined action
$$W(H_{\bar z\bar z}, {\cal U}_{z})=S (H_{\bar z\bar z})+S({\cal
U}_{ z})-\int d2zd2\theta H_{\bar z\bar z}{\cal
U}_{z}.\eqn\nabokhas$$ It can be easily checked that this combined
action is invariant under the transformations given by equations
\wafic\ and \sabra.

We turn now to find a solution for the action
$S(H_{\bar z\bar z})$.
Parametrizing $H_{\bar z \bar z}$ as
$$H_{\bar z \bar z}={\partial_{\bar
z}f+\bar\psi\partial_{\bar z}\psi+
\psi\partial_{\bar
z}\bar\psi\over\partial_{z}f+\bar\psi\partial_{z}\psi+
\psi\partial_{z}\bar\psi},$$ then the
anomaly  equation \Gabo\ is solved by $$Z_{z}=-{k\over2}\Big({\cal
S}(\psi)\Big)=-{k\over2}\Big({\partial_z D_\theta\bar\psi\over
D_\theta\bar\psi}- {\partial_z  D_{\bar\theta}\psi\over
D_{\bar\theta}\psi}-{2\partial_z\psi\partial_z\bar\psi\over
D_\theta\bar\psi D_{\bar\theta}\psi}\Big),\eqn\check$$ where ${\cal
S}(\psi)$ is the N=2 super-Schwartzian derivative [\cohen], and the
action  $S(H_{\bar z\bar z})$ is given by,  $$S(H_{\bar z\bar
z})={k\over2}S_{s.grav}{(2,0)}(f,\psi,\bar\psi).\eqn\checkagain$$
Obviously, the action of $S({\cal U}_{z})$ describes the geometric
quantization of the N=2 superVirasoro algebra and is given by the
action constructed on the coadjoint orbit of purely central
extension of N=2 superVirasoro group,
$$S({\cal U}_{z})=-{k\over2}S_{s.vir}{(2,0)}=-{k\over2}\int
d2zd2\theta\Big({\partial_z\Theta\partial_{\bar z}\bar\Theta-
\partial_z\bar\Theta\partial_{\bar z}\Theta\over(
D_{\bar\theta}\Theta)(D_{\theta}\bar\Theta)}\Big),\eqn\napalm$$
where$$\eqalign{L_{\bar z\bar z}=&{\partial_{\bar
z}X+\bar\Theta\partial_{\bar z}\Theta+ \Theta\partial_{\bar
z}\bar\Theta\over\partial_{z}X+\bar\Theta\partial_{z}\Theta+
\Theta\partial_{z}\bar\Theta},\cr {\cal
U}_{z}=&-{k\over2}\Big({\partial_z D_\theta\bar\Theta\over
D_\theta\bar\Theta}-{\partial_z D_{\bar\theta}\Theta\over
D_{\bar\theta}\Theta}-{2\partial_z\Theta\partial_z\bar\Theta\over
D_\theta\bar\Theta D_{\bar\theta}\Theta}\Big)=-{k\over2}{\cal
S}(\Theta).\cr}\eqn\jenny$$
The action \napalm\ is the reduced $(1,0)$
Osp(2,2) supergauge action. The original supergauge theory is
described by a $(1,0)$ WZNW model which has an N=1 Osp(2,2)
current algebra in its left-moving sector, while its right-moving
sector is described by a bosonic Osp(2,2) current algebra. In the
reduced supergauge theory, the symmetry of the left-moving part
is reduced to the N=2 superconformal symmetry, while the
right-moving sector still has the current algebra symmetry.

The finite form of \wafic\ and \sabra\
can be represented, respectively,  by  $$\eqalign{X(z,\bar
z,\theta,\bar\theta)&\rightarrow X(X_1,\bar
z,\Theta_1,\bar\Theta_1),\cr \Theta(z,\bar
z,\theta,\bar\theta)&\rightarrow \Theta(X_1,\bar
z,\Theta_1,\bar\Theta_1),\cr \bar\Theta(z,\bar
z,\theta,\bar\theta)&\rightarrow \bar\Theta(X_1,\bar
z,\Theta_1,\bar\Theta_1),\cr f(z,\bar
z,\theta,\bar\theta)&\rightarrow f(X_1,\bar
z,\Theta_1,\bar\Theta_1),\cr \psi(z,\bar
z,\theta,\bar\theta)&\rightarrow \psi(X_1,\bar
z,\Theta_1,\bar\Theta_1),\cr \bar\psi(z,\bar
z,\theta,\bar\theta)&\rightarrow \bar\psi(X_1,\bar
z,\Theta_1,\bar\Theta_1),\cr}\eqn\dallas$$ which for convienience
will be written as,   $$\eqalign{(X, \Theta, \bar\Theta)&\rightarrow
(X, \Theta,\bar\Theta)\bullet(X_1,\Theta_1,\bar\Theta_1),\cr
(f,\psi,\bar\psi)&\rightarrow
(f,\psi,\bar\psi)\bullet(X_1,\Theta_1,\bar\Theta_1).\cr}\eqn\hit$$
Using this notation, $(X{-1},\Theta{-1},\bar\Theta{-1})$ is
defined as,
$$(X,\Theta,\bar\Theta)\bullet(X{-1},\Theta{-1},\bar\Theta{-1})=
(z,\theta,\bar\theta).\eqn\susi$$ The invariance of the combined
action $W(H_{\bar z,\bar z}, {\cal U}_{z})$ under
superdiffeomorphisms implies the relationship, $$W\Big((X,
\Theta,\bar\Theta),(f,\psi,\bar\psi)\Big)= W\Big((X,
\Theta,\bar\Theta)\bullet(X_1,\Theta_1,\bar\Theta_{1}),
(f,\psi,\bar\psi)\bullet(X_1,\Theta_1,\bar\Theta_{1})\Big).
\eqn\joy$$ If we set
$(X_1,\Theta_1,\bar\Theta_{1})=(f{-1},\psi{-1},\bar\psi{-1})$ or
$(X_1,\Theta_1,\bar\Theta_1)=(X{-1},\Theta{-1},\bar\Theta{-1})$ ,
then the above equation gives
$$\eqalign{W\Big((X,\Theta,\bar\Theta),(f,\psi,\bar\psi)\Big)&=
-{k\over2}S_{s.vir}{(2,0)}\Big((X,\Theta,\bar\Theta)\bullet
(f{-1},\psi{-1},\bar\psi{-1})\Big)\cr
&={k\over2}S_{s.grav}{(2,0)}\Big((f,\psi,\bar\psi)\bullet
(X{-1},\Theta{-1},\bar\Theta{-1})\Big),\cr}\eqn\napper$$ and in
particular, $$-{k\over2}S_{s.vir}{(2,0)}\Big((X,\Theta,\bar\Theta
)\Big)
={k\over2}S_{s.grav}{(2,0)}\Big((X{-1},\Theta{-1},\bar\Theta{-1})\Big).
\eqn\sophie$$ This relation is the supergravitational analogue of
Eq.\israel,$$S_1(h{-1})=S_2(h)$$ and explains why one obtains the
induced $(2,0)$ $2d$ supergravity action from the geometric action
when
$$(X,\Theta,\bar\Theta)\bullet(f,\psi,\bar\psi)=(z,\theta,\bar\theta).\eqn\tel$$
Finally, we obtain the supergravitational composition formula
$$\eqalign{&S_{s.vir}{(2,0)}\Big((X,\Theta,\bar\Theta)\bullet
(f{-1},\psi{-1},\bar\psi{-1})\Big)=
S_{s.vir}{(2,0)}\Big((X,\Theta,\bar\Theta)\Big)\cr &+
S_{s.vir}{(2,0)}\Big((f{-1},\psi{-1},\bar\psi{-1})\Big) -\int
d2zd2\theta\Big({\partial_{\bar z}f+\bar\psi\partial_{\bar z}\psi+
\psi\partial_{\bar
z}\bar\psi\over\partial_{z}f+\bar\psi\partial_{z}\psi+
\psi\partial_{z}\bar\psi}\Big){\cal S}(\Theta).\cr}\eqn\oblivious$$
This is the $(2,0)$ supergravitational analogue of the composition
formula \police\ of the
$(1,0)$ super WZNW
model.

Let us summarize what we have done. The relationship existing
between the induced $(2,0)$ $2d$ supergravity in the superchiral
gauge formulation to the geometric action describing N=2
superconformal group is verified. Also, the action
$S_{s.vir}{(2,0)}$ is derived as a constrained supergauge theory.
The constrained theory has a  left-moving section  with the N=2
superconformal symmetry, while the right-moving part has an Osp(2,2)
current algebra. Thus our analysis provides a canonical derivation
of the Osp(2,2) current algebra of the induced $(2,0)$ $2d$
supergravity theory in the superchiral gauge.  \section{Discussion}
The analysis of this paper can be easily generalized to the case
of $(1,1)$ Osp(2,2) supergauge theory. In this case, the original
theory has both its left and right-moving sectors described by an
N=1 Osp(2,2) current algebras. Imposing the conditions \sod\ on
the left-moving sector gives a reduced theory with
the left-moving sector having the N=2 superconformal symmetry
and the right-moving sector having the N=1  Osp(2,2) current algebra.
The reduced action can be simply deduced by replacing
$\partial_{\bar z}$ by $D_{\theta-}$ in the action \napalm\
$$S={k\over2}\int d2zd2\theta d{\theta-}
\Big({\partial_z\Theta D_{\theta-}
\bar\Theta- \partial_z\bar\Theta D_{\theta-}\Theta\over(
D_{\bar\theta}\Theta)(D_{\theta}\bar\Theta)}\Big),\eqn\nlm$$
where $D_{\theta-}=\partial_{\theta-}+{\theta-}\partial_{\bar
z},$ $\theta-$ is a right-moving Grassmann coordinate and
$\Theta=\Theta(z,\bar z,\theta,\bar\theta,{\theta-})$ satisfies the
conditions \hell. We suggest that the action \nlm\ could be relevant
to the case of induced $(2,2)$ $2d$ supergravity. However the
superchiral formulation of this case and the derivation of the
associated current algebra remains to be analysed.

In the introduction it was pointed out that a
similar analysis to the one in [\diff] has been done in [\oog,\Alek]
using the Hamiltonian reduction method. In this method, one imposes a
constraint on the phase space of the SL(2,R) WZNW model, then the
constrained model has a residual symmetry which is then used to
gauge away a further degree of freedom giving a model describing the
geometric action of Virasoro algebra. The same method was also used
in [\supoog] where it was shown that the N=1 and N=2 superconformal
algebra has a hidden Osp(1,2) and Osp(2,2) current algebra
respectively. However, our calculations together with the analysis
of [\superwaf] suggests that the symplectic structure of N=1, 2
superconformal algebras can also be obtained via the Hamiltonian
reduction from that of N=1 Osp(1,2), Osp(2,2) current algebra
respectively. A complete analysis of this suggestion based on the
free field realization [\waki,\unpublished] of the super current
algebras will be reported on in a separate publication.

\centerline{ ACKNOWLEDGEMENT}

I would like to thank  C. Hull for useful
conversations.
\refout
\bye